\documentclass[preprint,preprintnumbers,amsmath,amssymb,
superscriptaddress,floatfix]{revtex4}
\usepackage{graphicx}
\usepackage{amsmath}
\usepackage{bm}
\usepackage{mathrsfs}
\usepackage{color}
\usepackage{slashed}
\usepackage{dcolumn}
\usepackage[normalem]{ulem}

\newcommand{\be}{\begin{equation}}
\newcommand{\ee}{\end{equation}}
\newcommand{\ba}{\begin{eqnarray}}
\newcommand{\ea}{\end{eqnarray}}
\newcommand{\nn}{\nonumber}

\begin{document}

\title{Study on Pentaqaurks by Solving Schrodinger Equation in the Non-Hermitian Quantum Mechanics}

\author{Bao-Xi Sun}
\email{sunbx@bjut.edu.cn}
\affiliation{School of Physics and Optoelectronic Engineering, Beijing University of Technology, Beijing 100124, China}


\begin{abstract}
The interaction of the charmed baryon and the anticharmed meson is assumed to be realized by exchanging a scalar meson of $f_0(500)$, and then these systems are studied by solving the Schrodinger equation, respectively. When the pentaquarks $P_{c\bar{c}}(4312)^{+}$, $P_{c\bar{c}}(4457)^{+}$, $P_{c\bar{c}s}(4338)^0$, and $P_{c\bar{c}s}(4459)^{0}$ are treated as $\Sigma_c \bar{D}$, $\Sigma_c \bar{D}^*$, $\Xi_c \bar{D}$ and $\Xi_c \bar{D}^*$ bound states, four resonance states of them are obtained as solutions of the Schrodinger equation under the outgoing wave condition, respectively. The resonance state of $\Sigma_c \bar{D}$ might correspond to the particle $P_{c\bar{c}}(4440)^{+}$, while the other three resonance states have no counterparts in the review of the Particle Data Group(PDG). Although the binding energy of the bound state is only several MeVs, all these resonance states are more than 100 MeV higher than their corresponding thresholds, respectively. The calculation results indicate that the spectrums of the strange and non-strange pentaquarks are symmetric to each other. 
\end{abstract}


\maketitle

\section{Introduction}
\label{Introduction}

The pentaquark state was predicted by Murry Gell-Mann in 1964 when his famous article on the quark model is published\cite{Gell-Mann:1964ewy}, and then many research works on pentaquarks have been done theoretically and experimentally. Especially, the $\Theta(1540)^+$ particle was announced to be a candidate of the pentaquark state and soon was denied\cite{ParticleDataGroup:2006fqo}.
Approximately ten years ago, the LHCb collaboration reported a significant $J/\psi p$ structure in the decay of $\Lambda^0_b \rightarrow J/\psi p K^{-}$\cite{LHCb:2015yax}, and an exotic structure with a quark content of $uudc\bar{c}$ was demonstrated. In 2019. Three peaks were determined at 4312, 4440 and 4457 MeV in the $J/\psi p$ mass, which were labeled $P_{c\bar{c}}(4312)^{+}$, $P_{c\bar{c}}(4440)^{+}$, and $P_{c\bar{c}}(4457)^{+}$, respectively\cite{LHCb:2019kea}. 
LHCb Collaboration observed a strange pentaquark state $P_{c\bar{c}s}(4459)^0$ in the process of $\Xi_b^- \rightarrow J/\psi \Lambda K^-$, which can decay into a $J/\psi$ particle and a $\Lambda$ hyperon\cite{LHCb:2020jpq}. Recently, another highly significant strange pentaquark state $P_{c\bar{c}s}(4338)^0$ was announced in the analysis of the decay of $B^- \rightarrow J/\psi \Lambda \bar{p}$\cite{LHCb:2022ogu}.
Ever since the discovery of pentaquarks in 2015, more attentions have been paid to the pentaquarks discovered by the LHCb Collaboration, and many analyzes have been done with different methods, such as the coupled-channel approximation\cite{Wu:2010jy,Chen:2016ryt,Shen:2019evi,Xiao:2019aya,Xiao:2019gjd,Liu:2019tjn,Wang:2019spc,Xiao:2020frg,Yalikun:2021bfm}, the chiral effecitve field theory\cite{Wang:2019nvm}, the QCD sum rule\cite{Liu:2019zoy,Wang:2022neq}, the Gursey-Radicati inspired mass formula\cite{Santopinto:2016pkp}, and the color-magnetic interaction model\cite{Li:2023aui}. More research wroks on pentaquarks can be found in the review article written by M. Karliner and T. Skwarnicki\cite{ParticleDataGroup:2024cfk}.

Since the $P_{c\bar{c}}(4312)^{+}$ particle is only $5.59$MeV below the $\Sigma_c^{+} \bar{D}^0$ threshold, it can be regarded as a $\Sigma_c^{+} \bar{D}^0$ bound state.
Similarly, the $P_{c\bar{c}}(4457)^{+}$ particle is approximately $2.2$MeV lower than the $\Sigma_c^{+} \bar{D}^{*0}$ threshold; therefore, it is more likely to be a
bound state of $\Sigma_c^{+} \bar{D}^{*0}$. Along this clue, we hope to evaluate the structure of these pentaquark states by solving the Schrodinger equation just as done in \cite{Sun:2024ezp}, where the $K \bar{K}^*$ and $D \bar{D}^*$ systems are studied with the assumption of one-pion exchange interaction. Of course, when the $\Sigma_c^{+} \bar{D}^0$ or $\Sigma_c^{+} \bar{D}^{*0}$ interaction is studied, instead of the pion, a scalar meson is assumed to play the role of intermediate meson between the charmed baryon and the anticharmed meson. This method is also extended to study the properties of strange pentaquarks.

The whole article is organized as follows. In Section~\ref{sect:framework}, the framework is evaluated in detail. 
The $\Sigma_c \bar{D}^0$, $\Sigma_c \bar{D}^*$, $\Xi_c \bar{D}$ and $\Xi_c \bar{D}^*$ systems are analyzed in Sections~\ref{SigmacplusbarD0}, \ref{SigmacplusbarDstar}, \ref{sect:XicbarD} and \ref{XicDstar}, respectively. Finally, a summary is given in Section~\ref{sect:summary}.

\section{The Schrodinger equation with a Yukawa-type potential}
\label{sect:framework}

The pentaquarks observed by LHCb Collaboration could be molecular states consisting of a charmed baryon and an anticharmed meson, and two-pion exchange plays a critical role in the formation of these molecular states\cite{Karliner2019}. At this point, it is similar to the interaction of the nucleon and the quarkonium in~\cite{TarrusCastella:2018php}, where a Yukawa-type potential is evaluated when the on-shell approximation is taken into account.
The scalar meson $f_0(500)$ can be treated as a resonance state of two pions; therefore, it is reasonable to assume that the two-pion exchange mode between the charmed baryon and the anticharmed meson can be replaced approximately with an intermediate scalar meson. In this situation, the interaction between the charmed baryon and the anticharmed meson takes a Yukawa-type potential,  
\be
\label{eq:202307071816}
V(r)=-g^2\frac{e^{-mr}}{d},
\ee
where $m$ is the mass of the scalar meson, $g$ is the coupling constant, and the distance $r$ in the denominator has been replaced with the force range $d=1/m$ approximately.
It is apparent that the potential in Eq.~(\ref{eq:202307071816}) is reasonable in the force range, and it is equal to the original Yukawa potential asymptotically at infinity. Under this approximation, the Schrodinger equation can be solved analytically.

Supposing the radial wave function $R(r)=\frac{u(r)}{r}$, the radial Schrodinger equation with $l=0$ can be written as
\be
\label{eq:202307081218}
-\frac{\hbar^2}{2\mu} \frac{d^2 u(r)}{dr^2}+V(r)u(r)=Eu(r),
\ee
where $\mu$ is the reduced mass of the two-body system.

With variable substitution
\be
r \rightarrow x=\alpha e^{-\beta r},~~~~0 \le x \le \alpha,
\ee
and
\be
u(r)=J(x),
\ee
the radial Schrodinger equation in Eq.~(\ref{eq:202307081218}) becomes
\be
\frac{d^2 J(x)}{d x^2}+\frac{1}{x} \frac{d J(x)}{d x} +\left[\frac{2\mu g^2}{d \beta^2} \frac{x}{\alpha}^{\frac{1}{d \beta}} \frac{1}{x^2}+\frac{2 \mu E}{\beta^2} \frac{1}{x^2} \right] J(x)=0.
\ee
Supposing
\be
\alpha=2g \sqrt{2 \mu d},~~~~\beta=\frac{1}{2d},
\ee
and
\be
\label{eq:rhoenergy}
\rho^2=-8d^2 \mu E,~~~~E \le 0,
\ee
the radial Schrodinger equation becomes the $\rho$th order Bessel equation, 
\be
\label{eq:202307081903}
\frac{d^2 J(x)}{d x^2}+\frac{1}{x} \frac{d J(x)}{d x} +\left[1-\frac{\rho^2}{x^2}\right] J(x)=0,
\ee
and its solution is the $\rho$th order Bessel function $J_\rho(x)$.

For the bound state, when $r \rightarrow +\infty$, the radial wave function $R(r) \rightarrow 0$, which implies that $u(r)=J_\rho(\alpha e^{-\beta r})=J_\rho(0)$ with $\rho \ge 0$. On the other hand, when $r \rightarrow 0$, $u(r) \rightarrow 0$, which means
\be
\label{eq:Besselzeropoint}
J_\rho(\alpha)=0.
\ee
Therefore, if only one bound state of the two-body system has been detected and the binding energy is given, the order of the Bessel function $\rho$ in Eq.~(\ref{eq:Besselzeropoint}) can be determined according to Eq.~(\ref{eq:rhoenergy}), and then the coupling constant $g$ in the Yukawa potential is obtained with the first nonzero zero point of the Bessel function $J_\rho(\alpha)$, which takes a form of
\be
\label{eq:couplingzeropoint}
g^2=\frac{\alpha^2}{8 \mu  d}.
\ee

The second kind of Hankel function $H_\rho^{(2)}(x)$ is also an independent solution of the Bessel equation. In reality,  
$H_\rho^{(2)}(x)$ represents a wave leaving from the coordinate origin $r=0$. When $r\rightarrow+\infty$,
$u(r) \sim H_\rho^{(2)}(\alpha e^{-\beta r})\rightarrow H_\rho^{(2)}(0)$. When $r \rightarrow 0$, $u(0) \rightarrow 0$, which implies
\be
\label{eq:outgoing}
H_\rho^{(2)}(\alpha)=0.
\ee
If the first nonzero zero point of the Bessel function $J_\rho(\alpha)$ has been obtained according to the binding energy of the two-body system, which indicates the coupling constant $g$ in the Yukawa-type potential is determined, the energy of the two-body resonance state can be calculated with the outgoing wave condition in Eq.~(\ref{eq:outgoing}). Apparently, the energy of the resonance state is related to the order of the second kind of Hankel function and takes a complex value, 
$E=M-i\frac{\Gamma}{2}$, where the real part represents the mass of the resonance state, and the imaginary part is one-half of the decay width, that is, $\Gamma=-2iImE$, as discussed in Ref.~\cite{Moiseyev}.

\section{$\Sigma^+_c \bar{D}^0$ and $\Sigma^{++}_c D^-$ interactions}
\label{SigmacplusbarD0}

\begin{figure}
\includegraphics[width=0.8\textwidth]{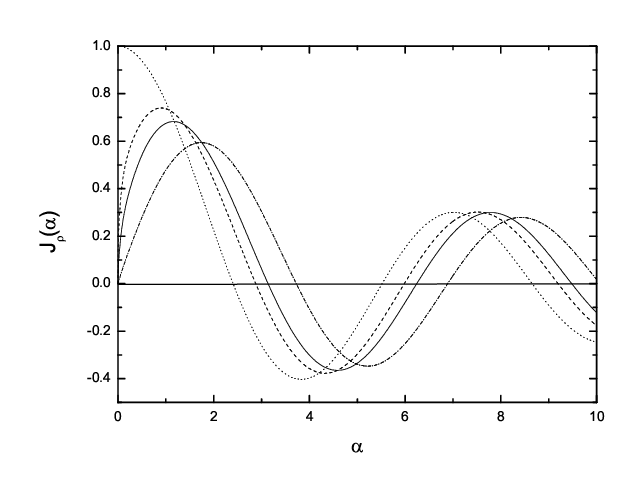}
\caption{Bessel functions $J_\rho(\alpha)$ with $\rho=0$ and the first zero point at $\alpha=2.405$ for the channel of $\Xi_c^{+} D^-$(Dot line), $\rho=0.317$ and the first nonzero zero point at $\alpha=2.88$ for the channel of $\Sigma_c^{+} \bar{D}^{*0}$(Dash line), $\rho=0.495$ and the first nonzero zero point at $\alpha=3.14$ for the channel of $\Sigma_c^{+} \bar{D}^0$(Solid line) and $\rho=0.920$ and the first nonzero zero point at $\alpha=3.73$ for the channel of $\Xi_c^0 \bar{D}^{*0}$(Dash-dot line) are depicted from left to right, respectively.}
\label{fig:zeropoint}
\end{figure}

\begin{figure}
\includegraphics[width=0.8\textwidth]{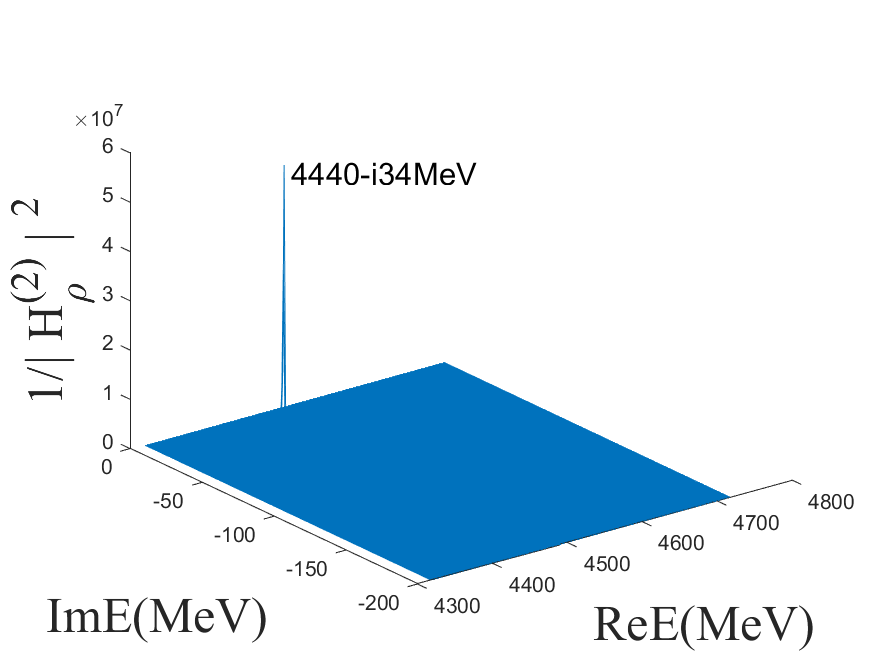}
\caption{$1/|H_\rho^{(2)}(\alpha)|^2$ .vs. the complex energy $E$ with $\alpha=3.14$ and $m_\sigma=440$ MeV in the $\Sigma_c^+ \bar{D}^{0}$ case. The pole of $1/|H_\rho^{(2)}(\alpha)|^2$ corresponds to a zero point of the second kind of Hankel function $H_\rho^{(2)}(\alpha)$, which represents a resonance state of $\Sigma_c^+ \bar{D}^{0}$, as labeled in the figure.}
\label{fig:D0Sigcp}
\end{figure}

It is assumed that a scalar meson is exchanged between the $\Sigma^+_c$ baryon and the $\bar{D}^0$ meson, and then the $\Sigma^+_c \bar{D}^0$ interaction is investigated by solving the Schrodinger equation. 
Suppose that the particle $P_{c\bar{c}}(4312)^{+}$ is a bound state of $\Sigma^+_c \bar{D}^0$ with a binding energy of 5.59 MeV, the first nonzero zero point of the Bessel function $J_\rho(\alpha)$ lies at $\alpha=3.14$ according to Eqs. (\ref{eq:rhoenergy}) and (\ref{eq:Besselzeropoint}), which is depicted in Fig.~\ref{fig:zeropoint}. Under the outgoing wave condition in Eq.~(\ref{eq:outgoing}), a $\Sigma^+_c \bar{D}^0$ resonance state is generated dynamically at $4440-i34$ MeV in the complex energy plane when the mass of the scalar meson takes a value of 440 MeV, as shown in Fig.~\ref{fig:D0Sigcp}.
Similarly, if the particle $P_{c\bar{c}}(4312)^{+}$ is treated as a $\Sigma^{++}_c D^-$ bound state, the binding energy would be 11.57 MeV. In this case, the first nonzero zero point of the Bessel function takes a value of $\alpha=3.44$, and a resonance state of $\Sigma^{++}_c D^-$ appears at $4445-i64$ with $m_\sigma=440$ MeV.  
Therefore, it is reasonable to assume that the particle $P_{c\bar{c}}(4440)^{+}$, which is reported by the LHCb Collaboration, might be a resonance state of $\Sigma^+_c \bar{D}^0$ or $\Sigma^{++}_c D^-$.

Although the decay widths of the pentaquarks are all narrow, we still think that it is necessary to take into account the influence of the decay widths when the zero point of the Bessel function is determined. Therefore, the binding energy would be complex when the particle $P_{c\bar{c}}(4312)^{+}$  is treated as a bound state of $\Sigma^+_c \bar{D}^0$ or $\Sigma^{++}_c D^-$, respectively.  
For the case of $\Sigma^+_c \bar{D}^0$, the binding energy takes a value of $5.59+i4.9$ MeV, and the first nonzero zero point of the Bessel function is $\alpha=3.29+i0.32$ with the mass of the scalar meson $m_\sigma=390$ MeV. Sequently, a resonance state of $\Sigma^+_c \bar{D}^0$ is detected at $4440-i38$ MeV in the complex energy plane, which is relevant to the order of the second Hankel function $H_\rho^{(2)}(\alpha)$ with $\alpha=3.29+i0.32$.
Similarly, if the particle $P_{c\bar{c}}(4312)^{+}$ is a bound state of $\Sigma^{++}_c D^-$, the binding energy is $11.57+i4.9$ MeV and the first nonzero zero point of the corresponding Bessel function is $\alpha=3.60+i0.23$ with $m_\sigma=390$ MeV. a resonance state of $\Sigma^{++}_c D^-$ is found at $4437-i64$ MeV when the outgoing wave condition in Eq.~(\ref{eq:outgoing}) is taken into account.
Since the Schrodinger equation is solved in the S-wave approximation, the spin and parity of the particle $P_{c\bar{c}}(4440)^{+}$ can be determined as $J^P=\frac{1}{2}^-$.
The calculation results of $\Sigma^+_c \bar{D}^0$ and $\Sigma^{++}_c D^-$ interactions are summarized in Table~\ref{table:DSigc} .

In conclusion, the exchange of the scalar meson plays a critical role when the interactions of $\Sigma^+_c \bar{D}^0$ and $\Sigma^{++}_c D^-$ are considered. The particle $P_{c\bar{c}}(4312)^{+}$ might be a bound state of $\Sigma^+_c \bar{D}^0$ or $\Sigma^{++}_c D^-$, while the particle $P_{c\bar{c}}(4440)^{+}$ would be a resonance state of them when the Schrodinger equation is solved under the outgoing wave condition.

\begin{table}[htbp]
 \renewcommand{\arraystretch}{1.2}
\centering
\vspace{0.5cm}
\begin{tabular}{c|c|c|c|c|c|c|c|c}
\hline\hline
Channel& $B$ & $m_\sigma$  & $\alpha$ & $E$ &   Name & $I^G(J^{PC})$ &  $M$ & $\Gamma$ \\
 \hline
$\Sigma_c^{+} \bar{D}^0$& $5.59$  & 440   &$3.14$ & $4440-i34$ & $P_{c\bar{c}}(4440)^+$ & $\frac{1}{2}^+(?^{?})$ &   $4440.3\pm1.3^{+4.1}_{-4.7}$  & $20.6\pm4.9^{+8.7}_{-10.1}$
  \\
& $5.59+i4.9$ & 390 & $3.29+i0.32$ & $4440-i38$ &  &  &     & 
  \\  
\hline
$\Sigma_c^{++} D^-$ & $11.57$ & 440   &$3.44$ & $4445-i64$ &  &  &     &   \\
& $11.57+i4.9$ & 390 & $3.60+i0.23$ & $4437-i64$ &  &  &     & 
  \\  
\hline\hline
\end{tabular}
\caption{The complex energy $E$ of the possible $\Sigma_c^{+} \bar{D}^0$ ($\Sigma_c^{++} D^-$) resonance state with two schemes and the properties of the possible counterpart $P_{c\bar{c}}(4440)^+$ are displayed, while $B$ represents the binding energy of the particle $P_{c\bar{c}}(4312)^{+}$ as a bound state of $\Sigma_c \bar{D}$. $B$, $m_\sigma$, $E$, $M$ and $\Gamma$ are all in units of MeV.}
\label{table:DSigc}
\end{table}

\section{$\Sigma^+_c \bar{D}^{*0}$ and $\Sigma^{++}_c D^{*-}$ interactions}
\label{SigmacplusbarDstar}

\begin{figure}
\includegraphics[width=0.8\textwidth]{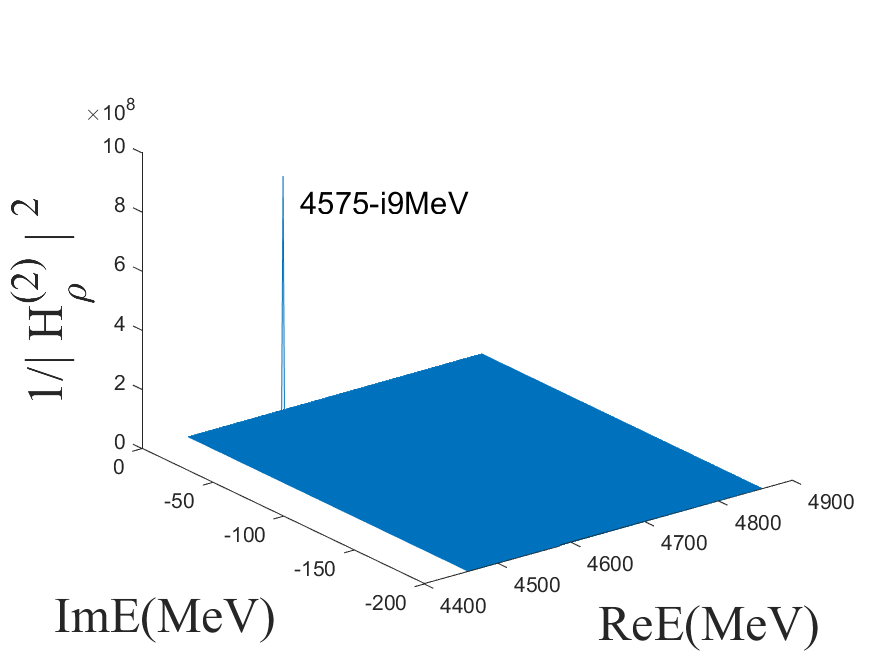}
\caption{$1/|H_\rho^{(2)}(\alpha)|^2$ .vs. the complex energy $E$ with $\alpha=2.88$ and $m_\sigma=440$ MeV in the $\Sigma_c^+ \bar{D}^{*0}$ case. The pole of $1/|H_\rho^{(2)}(\alpha)|^2$ corresponds to a zero point of the second kind of Hankel function $H_\rho^{(2)}(\alpha)$, which represents a resonance state of $\Sigma_c^+ \bar{D}^{*0}$, as labeled in the figure.}
\label{fig:Dstar0Sigcp}
\end{figure}

Since the particle $P_{c\bar{c}}(4457)^{+}$ is only 2.2 MeV below the threshold $\Sigma^+_c \bar{D}^{*0}$, it can be regarded as a bound state of $\Sigma^+_c \bar{D}^{*0}$. If the exchange of the scalar meson with the mass $m_\sigma=440$ MeV is assumed, the first nonzero zero point of the Bessel function appears at $\alpha=2.88$, as depicted in Fig.~\ref{fig:zeropoint}. When the outgoing wave condition in Eq.~(\ref{eq:outgoing}) is considered, a resonance state of $\Sigma^+_c \bar{D}^{*0}$ appears at $4575-i9$ MeV in the complex energy plane, as shown in Fig.~\ref{fig:Dstar0Sigcp}. 

If the particle $P_{c\bar{c}}(4457)^{+}$ is treated as a bound state of $\Sigma^{++}_c D^{*-}$, the binding energy would be $6.93$ MeV, the first nonzero zero point of the Bessel function is $\alpha=3.23$, and a resonance state of
$\Sigma^{++}_c D^{*-}$ is generated dynamically, which is located at $4582-i41$ MeV in the complex energy plane. 

If the decay width of the particle $P_{c\bar{c}}(4457)^{+}$ is taken into account, $\Gamma=6.4$ MeV, the first nonzero zero point of the Bessel function is a complex number, $\alpha=3.03+i0.32$ for the channel of $\Sigma^+_c \bar{D}^{*0}$, and $\alpha=3.36+i0.20$ for the channel of $\Sigma^{++}_c D^{*-}$ with the mass of the intermediate scalar meson $m_\sigma=390$ MeV. Therefore, under the outgoing wave condition in Eq.~(\ref{eq:outgoing}), a resonance state of $\Sigma^+_c \bar{D}^{*0}$ appears at $4576-i15$ MeV in the complex plane. For the $\Sigma^{++}_c D^{*-}$ channel, a resonance state is located at $4572-i42$ MeV, where the corresponding threshold energies have been included in these values. 
Apparently, if the particle $P_{c\bar{c}}(4457)^{+}$ is a bound state of $\Sigma^+_c \bar{D}^{*0}$ or $\Sigma^{++}_c D^{*-}$, there would be a corresponding resonance state of $\Sigma^+_c \bar{D}^{*0}$ or $\Sigma^{++}_c D^{*-}$, which is approximately 110 MeV above the threshold, consists of five quarks, $c\bar{c}uud$, and can be regarded as a pentaquark state with $J^P=\frac{1}{2}^-$ or $\frac{3}{2}^-$. The calculation results of $\Sigma^+_c \bar{D}^{*0}$ and $\Sigma^{++}_c D^{*-}$ interactions are summarized in Table~\ref{table:DstarSigc}.

\begin{table}[htbp]
 \renewcommand{\arraystretch}{1.2}
\centering
\vspace{0.5cm}
\begin{tabular}{c|c|c|c|c}
\hline\hline
Channel & $B$ & $m_\sigma$  & $\alpha$ & $E$  \\
 \hline
$\Sigma_c^{+} \bar{D}^{*0}$ & $2.2$ & 440   &$2.88$ & $4575-i9$   \\
& $2.2+i3.2$ & 390 & $3.03+i0.32$ & $4576-i15$   \\  
\hline
$\Sigma_c^{++} D^{*-}$ & $6.93$ & 440   &$3.23$ & $4582-i41$  \\
& $6.93+i3.2$ & 390 & $3.36+i0.20$ & $4572-i42$   \\  
\hline\hline
\end{tabular}
\caption{
The complex energy $E$ of the possible $\Sigma_c^{+} \bar{D}^{*0}$ ($\Sigma_c^{++} D^{*-}$) resonance state with two schemes, while $B$ represents the binding energy of the particle $P_{c\bar{c}}(4457)^{+}$ as a bound state of $\Sigma_c \bar{D}^*$. $B$, $m_\sigma$ and $E$ are all in units of MeV.
}\label{table:DstarSigc}
\end{table}

\section{$\Xi_c^+ D^-$ interaction}
\label{sect:XicbarD}

In the decay of $B^-\rightarrow J/\psi \Lambda \bar{p}$, a strange pentaquark state $P_{c\bar{c}s}(4338)^0\rightarrow J/\psi \Lambda$ is observed with high significancy($>14\sigma$)\cite{LHCb:2022ogu}. Its mass $4338.2\pm0.7\pm0.4$ MeV is right at the $\Xi_c^+ D^-$ threshold, its width is narrow $7.0\pm1.2\pm1.3$ MeV, and its spin is $1/2$ with a preference for negative parity, which all fits the molecular model very well. This makes the $P_{c\bar{c}s}(4338)^0$ state a likely $\Xi_c \bar{D}$ analog of the $P_{c\bar{c}}(4312)^+$ state in the $\Sigma_c \bar{D}$ system.

\begin{figure}
\includegraphics[width=0.8\textwidth]{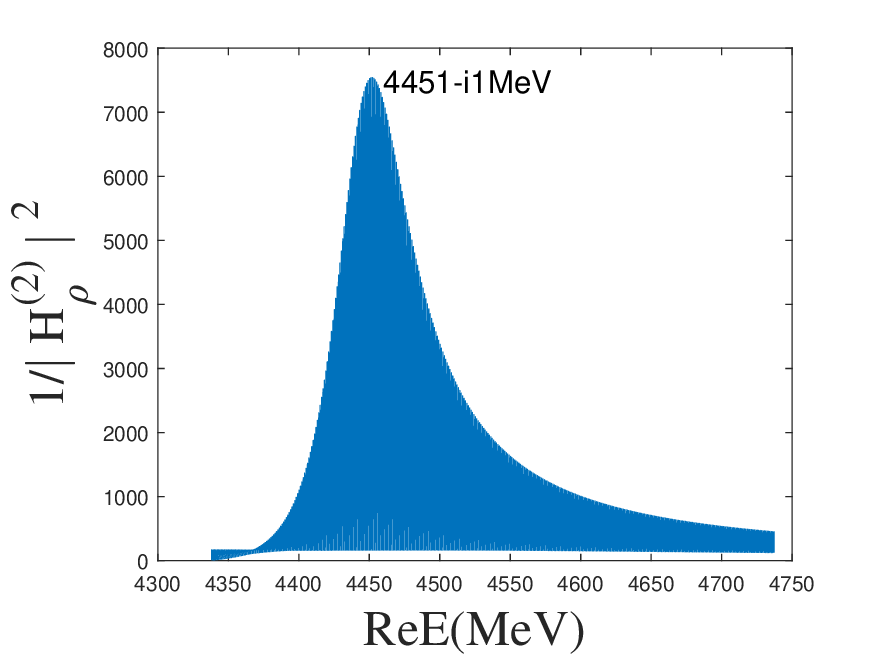}
\caption{$1/|H_\rho^{(2)}(\alpha)|^2$ .vs. the real part of the complex energy $E$ with $\alpha=2.405$ and $m_\sigma=440$ MeV in the $\Xi_c^+ {D}^{-}$ case. The pole of $1/|H_\rho^{(2)}(\alpha)|^2$ corresponds to a zero point of the second kind of Hankel function $H_\rho^{(2)}(\alpha)$, which represents a resonance state of $\Xi_c^+ {D}^{-}$, as labeled in the figure.}
\label{fig:D0Xic02405}
\end{figure}

The $P_{c\bar{c}s}(4338)^0$ is located just at the $\Xi_c^+ D^-$ threshold, which is similar to the case of the pentaquark $\chi_{c1}(3872)$, which lies at the $D\bar{D}^*$ threshold.
Similarly to the study in Ref.~\cite{Sun:2024ezp}, the $P_{c\bar{c}s}(4338)^0$ can be treated as a bound state of $\Xi_c^+ D^-$ with a binding energy of zero. Therefore, according to Eqs.~(\ref{eq:rhoenergy}), (\ref{eq:Besselzeropoint}) and (\ref{eq:couplingzeropoint}), the coupling constant of the $\Xi_c^+ D^-$ interaction can be calculated with the first zero point of the zeroth Bessel function, which is $g=0.55$ approximately.
With the same zero point $\alpha=2.405$, a resonance state of $\Xi_c^+ D^-$ with $J^P=\frac{1}{2}^-$ or $\frac{3}{2}^-$ can be evaluated according to the outgoing wave condition in Eq.~(\ref{eq:outgoing}), which lies at $4451-i1$ MeV in the complex energy plane, and is approximately 115 MeV above the threshold of $\Xi_c^+ D^-$, as depicted in Fig.~\ref{fig:D0Xic02405}. The calculation result of the $\Xi_c^+ D^-$ interaction is listed in Table~\ref{table:DXic}.

\begin{table}[htbp]
 \renewcommand{\arraystretch}{1.2}
\centering
\vspace{0.5cm}
\begin{tabular}{c|c|c|c|c}
\hline\hline
Channel & $B$ &$m_\sigma$  & $\alpha$ & $E$  \\
 \hline
$\Xi_c^{+} D^-$ & 0 & 440   &$2.405$ & $4451-i1$   \\
\hline\hline
\end{tabular}
\caption{The complex energy $E$ of the possible $\Xi_c^{+} D^-$ resonance state is displayed, while $B$ represents the binding energy of the particle $P_{c\bar{c}s}(4338)^0$ as a bound state of $\Xi_c^{+} D^-$. $B$, $m_\sigma$ and $E$ are  all in units of MeV.}
\label{table:DXic}
\end{table}

\section{$\Xi^0_c \bar{D}^{*0}$ and $\Xi^{+}_c D^{*-}$ interactions}
\label{XicDstar}

\begin{figure}
\includegraphics[width=0.8\textwidth]{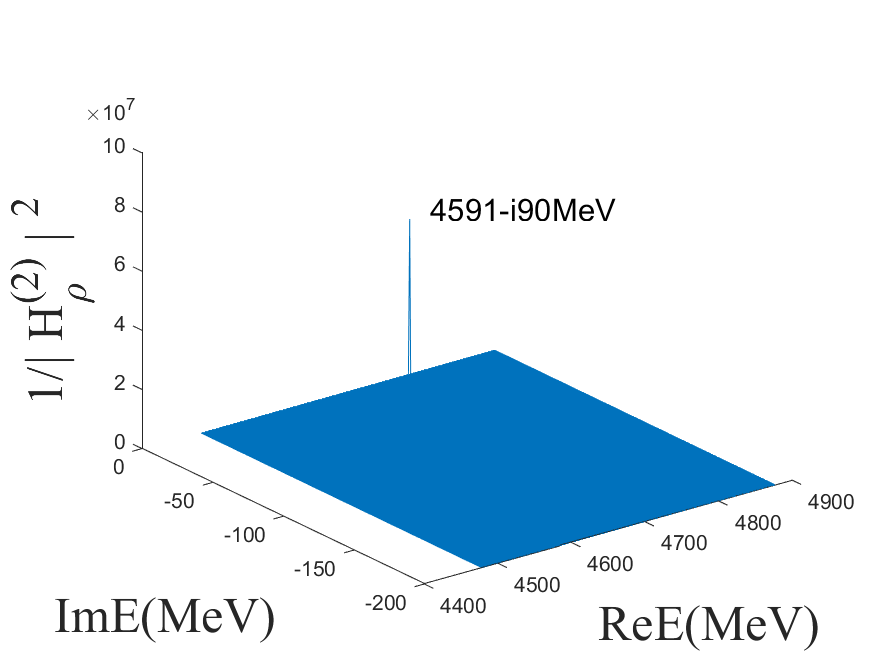}
\caption{$1/|H_\rho^{(2)}(\alpha)|^2$ .vs. the complex energy $E$ with $\alpha=3.73$ and $m_\sigma=440$ MeV in the $\Xi_c^0 \bar{D}^{*0}$ case. The pole of $1/|H_\rho^{(2)}(\alpha)|^2$ corresponds to a zero point of the second kind of Hankel function $H_\rho^{(2)}(\alpha)$, which represents a resonance state of $\Xi_c^0 \bar{D}^{*0}$, as labeled in the figure.}
\label{fig:Dstar0Xic0}
\end{figure}

The LHCb Collaboration observed a strange pentaquark $P_{c\bar{c}s}(4459)^0$ in the analysis of the decay process $\Xi_b^{-} \rightarrow J/\psi \Lambda K^-$\cite{LHCb:2020jpq}, 
whose mass and decay width were measured as
\ba
M&=&4458.8\pm2.9^{+4.7}_{-1.1}MeV, \\ \nn
\Gamma&=&17.3\pm6.5^{+8.0}_{-5.7}MeV,
\ea
respectively.
Since the mass of the particle $P_{c\bar{c}s}(4459)^0$ is $18.49(19.41)$ MeV below the threshold of $\Xi^0_c \bar{D}^{*0}(\Xi^{+}_c D^{*-})$. Therefore, it might be a bound state of $\Xi^0_c \bar{D}^{*0}$ or $\Xi^{+}_c D^{*-}$. In this case, the coupling constant in the Yukawa-type potential of $\Xi^0_c \bar{D}^{*0}$ or $\Xi^{+}_c D^{*-}$ can be determined if an intermediate scalar meson is exchanged between the baryon and the vector meson, which is related to the first non-zero zero point of the corresponding Bessel function. In the channel of $\Xi^0_c \bar{D}^{*0}$, the first nonzero zero point of the Bessel function appears at $\alpha=3.73$ with a mass of the scalar meson $m_\sigma=440$ MeV, as depicted in Fig.~\ref{fig:zeropoint}, and a resonance state of $\Xi^0_c \bar{D}^{*0}$ is detected at $4591-i90$ MeV in the complex energy plane when the outgoing wave condition is taken into account, which is shown in Fig.~\ref{fig:Dstar0Xic0}.
Analytically, in the channel of $\Xi^{+}_c D^{*-}$, the first nonzero zero point of the Bessel function is $\alpha=3.76$ with a mass of the scalar meson $m_\sigma=440$ MeV, and a resonance state of $\Xi^{+}_c D^{*-}$ appears almost at the same position in the complex energy plane under the outgoing wave condition in Eq.~(\ref{eq:outgoing}).

If the decay width of the strange pentaquark $P_{c\bar{c}s}(4459)^0$ is considered when the coupling constant in the interaction of $\Xi^0_c \bar{D}^{*0}$ or $\Xi^{+}_c D^{*-}$ is determined, the first nonzero zero point of the Bessel function takes a complex value, $\alpha=3.92+i0.31$ for the channel of $\Xi^0_c \bar{D}^{*0}$ and $\alpha=3.95+i0.30$ for the channel of $\Xi^{+}_c D^{*-}$. Eventually, when the outgoing wave condition is considered, a resonance state with $J^P=\frac{1}{2}^-$ or $\frac{3}{2}^-$ can be obtained at $4590-i91$ MeV in the channel of $\Xi^0_c \bar{D}^{*0}$, or $4589-i94$ MeV in the channel of $\Xi^{+}_c D^{*-}$, respectively. 
The calculation results of $\Xi^0_c \bar{D}^{*0}$ and $\Xi^{+}_c D^{*-}$ interactions are summarized in Table~\ref{table:DstarXic}.

The strange pentaquarks have been predicted in both molecular and compact pentaquark models\cite{Wu:2010jy,Shen:2019evi,Chen:2016ryt,Santopinto:2016pkp,Xiao:2019gjd,Wang:2019nvm}. Apparently, the number of possible pentaquark states is suppressed greatly when the baryon and the meson are assumed to be building blocks. In contrast, in the compact pentaquark model, two states with $J^P$ equal to $1/2^-$ and
$3/2^-$ are predicted, whose masses(widths) are $4454.9\pm2.7$ MeV $(\Gamma=7.5\pm9.7 MeV)$ and $4467.8\pm3.7$ MeV $(\Gamma=5.2\pm5.3 MeV)$, respectively. The hypothesis of two strange pentaquarks is also consistent with the experimental data of the LHCb Collaboration. 
However, in the method used in this work, where the Schrodinger equation is solved with a Yukawa-type potential of $\Xi^0_c \bar{D}^{*0}$ or $\Xi^{+}_c D^{*-}$, it is impossible that there exist two bound states of $\Xi_c \bar{D}^*$ because the attractive interaction of $\Xi^0_c \bar{D}^{*0}$ or $\Xi^{+}_c D^{*-}$ is too weak. Unless the value of the first nonzero zero point of the Bessel function is higher than 5.520, which is the value of the second zero point of the zeroth Bessel function, only one bound state of $\Xi_c \bar{D}^*$ exists.

\begin{table}[htbp]
 \renewcommand{\arraystretch}{1.2}
\centering
\vspace{0.5cm}
\begin{tabular}{c|c|c|c|c}
\hline\hline
Channel & $B$& $m_\sigma$  & $\alpha$ & $E$  \\
 \hline
$\Xi_c^{0} \bar{D}^{*0}$ & $18.49$ & 440   &$3.73$ & $4591-i90$   \\
                   & $18.49+i8.65$ & 390 & $3.92+i0.31$ & $4590-i91$   \\  
\hline
$\Xi_c^{+} D^{*-}$ & $19.41$ & 440   &$3.76$ & $4591-i93$  \\
                   & $19.41+i8.65$ & 390 & $3.95+i0.30$ & $4589-i94$   \\  
\hline\hline
\end{tabular}
\caption{The complex energy $E$ of the possible $\Xi_c^{0} \bar{D}^{*0}$ ($\Xi_c^{+} D^{*-}$) resonance state with two schemes, respectively, while $B$ represents the binding energy of the particle $P_{c\bar{c}s}(4459)^0$ as a bound state of $\Xi_c \bar{D}^{*}$. $B$, $m_\sigma$ and $E$ are all in units of MeV.}
\label{table:DstarXic}
\end{table}

\section{Summary}
\label{sect:summary}

The scalar meson is assumed to be exchanged between the charmed baryon and the anti-charmed meson. Since the scalar meson $f_0(500)$ can be regarded as a resonance state of two pions, it indicates that the two-pion exchange plays a critical role in the interaction of the charmed baryon and the anticharmed meson. In this proposition, it is consistent with the interaction of the nucleon and the quarkonium mentioned in Ref.~\cite{TarrusCastella:2018php}. 

The $\Sigma_c \bar{D}$, $\Sigma_c \bar{D}^*$, $\Xi_c \bar{D}$, and $\Xi_c \bar{D}^*$
systems are studied by solving the Schrodinger equation under different boundary conditions of the wave function.
The eigenenergy of the Hamiltonian can be complex when the outgoing wave condition is taken into account, and the complex solution of the Schrodinger equation might correspond to a resonance state of the interacting particles. 
Therefore, the generation of the resonance state can be regarded as a non-Hermitian problem in quantum mechanics.

By fitting the binding energy of the particle $P_{c\bar{c}}(4312)^{+}$, which is regarded as a bound state of $\Sigma_c \bar{D}$  in this work, the coupling constant in the $\Sigma_c \bar{D}$ interaction is determined.
Sequently, a resonance state of $\Sigma_c \bar{D}$ is generated as a solution of the Schrodinger equation under the outgoing wave condition, which appears in the vicinity of 4440 MeV in the complex energy plane and might correspond to the pentaquark $P_{c\bar{c}}(4440)^{+}$ in the PDG data.
Since the pentaquark $P_{c\bar{c}}(4457)^{+}$ is just several MeVs below the $\Sigma_c \bar{D}^*$ threshold, it can be considered as a bound state of $\Sigma_c \bar{D}^*$. Therefore, with the same method, a resonance state of $\Sigma_c \bar{D}^*$ is detected at 4575 MeV in the complex energy plane as a solution of the Schrodinger equation. 

This method is extended to study the properties of strange pentaquarks. 
Suppose that the $P_{c\bar{c}s}(4338)^0$ state is a $\Xi_c \bar{D}$ bound state with a binding energy of zero, a corresponding resonance state of $\Xi_c \bar{D}$ appears at 4451 MeV by solving the Schrodinger equation under the outgoing wave condition, whose spin and parity is $1/2$ and negative. Since the mass of the resonance state of $\Xi_c \bar{D}$ at 4451 MeV is less than that of $P_{c\bar{c}s}(4459)^0$, it is not possible to be a
counterpart of the particle $P_{c\bar{c}s}(4459)^0$.
Moreover, if the pentaquark $P_{c\bar{c}s}(4459)^0$ is treated as a bound state of $\Xi_c \bar{D}^*$, a resonance state of $\Xi_c \bar{D}^*$ is generated around 4590 MeV in the complex energy plane when the Schrodinger equation is solved under the outgoing wave condition. 

All resonance states of the charmed baryon and the anticharmed meson are more than 100 MeV higher than their corresponding thresholds. 
Especially, it is found that the spectrums of the pentaquark state with and without strange quarks are symmetric to each other, and the mass and width of them are analogous, respectively. It implies that
only non-strange meson is exchanged between the charmed baryon and the anticharmed meson, and the strange meson is forbidden. At this point, the assumption of the scalar-meson exchange is reasonable.
It is apparent that the elastic scattering process is dominant in the generation of the resonance states, where the initial particles are the same as those final ones.

The pentaquarks are treated as bound and resonance states of different charmed baryons and anticharmed mesons, and the resonance states of the charmed baryon and the anticharmed meson are generated dynamically as solutions of the Schrodinger equation when the outgoing wave condition is taken into account.
Although the bound states are only several MeVs below the threshold, 
all resonance states are more than 100 MeV higher than their corresponding thresholds.  At this point, it is different from the calculation results of the unitary coupled-channel approximation, where all the bound and resonance states appear only in the vicinity of the threshold. 

\begin{acknowledgments}
Bao-Xi Sun thanks Wren Yamada for useful discussions. 
\end{acknowledgments}


\end{document}